# Effect of Cold Hydrogen Plasma on Optical Parameters of an Optical Fiber

M.Medhat[a], S.Y.El-Zaiat[a], M.A.Saudy[a] and H.S.Hassanein[b]

[a] Physics Department, Faculty of Science, Ain Shams University, Abbasia, 11566, Cairo, Egypt;
[b] Basic Science Department, Faculty of Engineering, Misr University for Science and Technology,6th of October, 77, Giza, Egypt

## Abstract

Cold plasma of power 13W has been applied on step index plastic optical fibers. It is used to study the ability for any variations in the optical parameters that are effective in communication .The effect of plasma on the optical parameters, such as core index profile, cladding index and the numerical aperture, has been studied by multiple beam interference fringes. The interference fringes have been processed using a homemade Matlab written program to get more accurate measurements. It is found that there is an increase of 1.5839 in core index, at core-clad interface, with a plasma exposure time of 150 sec.

## 1. Introduction

Optical fibers are important tool to transmit signals quickly with more security and high accuracy of transmission. There are many classifications of optical fibers depending on: the fiber material (glass and polymer), way of signal propagation and the fiber diameter (single mode and multimode) and the relation between the fiber refractive index versus the fiber radius that called index profile (step index and graded index).

Glass fibers have many features that make it preferable to be used in long distances for hundreds of kilometers, but it suffers from difficulty of handling and need of long time for training, especially in case of cables connection[1]. Plastic optical fibers (POF) have more advantages such as low cost, large numerical aperture and suitable flexibility that give it priority for short -distances of hundreds of meters [2-3]. Many trials have been made to improve and increase the quality of signal transmission through POF [4-5].

Optical interferometry is one of the most important accurate methods for characterization of the fiber parameters [6-8]. In our work we use Fizeau interferometer that gives accurate results of measurements. Also, image processing and calculation process with Matlab is helpful to get accurate data from the interferogram that obtained from the fiber which is inserted in the wedge of the interferometer.

Plasma treatment is a field for processing of many materials and solving many problems especially in the textile field. Plasma, which is known as an ionized gas, consists of electrons, ions and neutral particles. These particles are atoms and molecules with high activity and are characterized by ionizing states that are thermodynamically stable [9]. Glow-discharge plasma gives excellent results in the surface modification of materials and industrial components [9-11]. Plasma processing has been widely used in a large number of technologies, from metallurgy to the manufacturing of computer chips, and covers a wide variety of materials, including metals, semiconductors, and polymers [12].

In our experiment, we have studied the effect of hydrogen cold plasma DC- discharge on step index plastic optical fiber. The fiber core is made of ploy methyl methacrylate (PMMA) and the cladding material is made from fluorinated polymer. Multiple beam interference fringes have been applied to detect the variation in some optical parameters associated with optical fibers. Fourier Transform Infrared (FTIR) spectroscopy has been applied on unexposed and exposed samples of the fiber to make a diagnostic interpretation of the changes in the chemical bonds of the samples after exposing to plasma [13].

# 2. Theoretical consideration

## 2.1. Optical path in circular fiber

The interferometric investigation is done by inserting the fiber in a silvered liquid wedge with the fiber axis is perpendicular to the edge of wedge .Figure 1 shows the cross section of the fiber inside liquid wedge interferometer. There are some parameters that must be known for step index optical fiber: radius of the fiber ($r_f$) ,radius of the core ($r_c$) ,refractive index of the cladding material ($n_{cl}$).and refractive index of the core material ($n_c$).A parallel mono chromatic beam of wave length λ is incident to AB and CD normal to the lower component of the interferometric wedge.

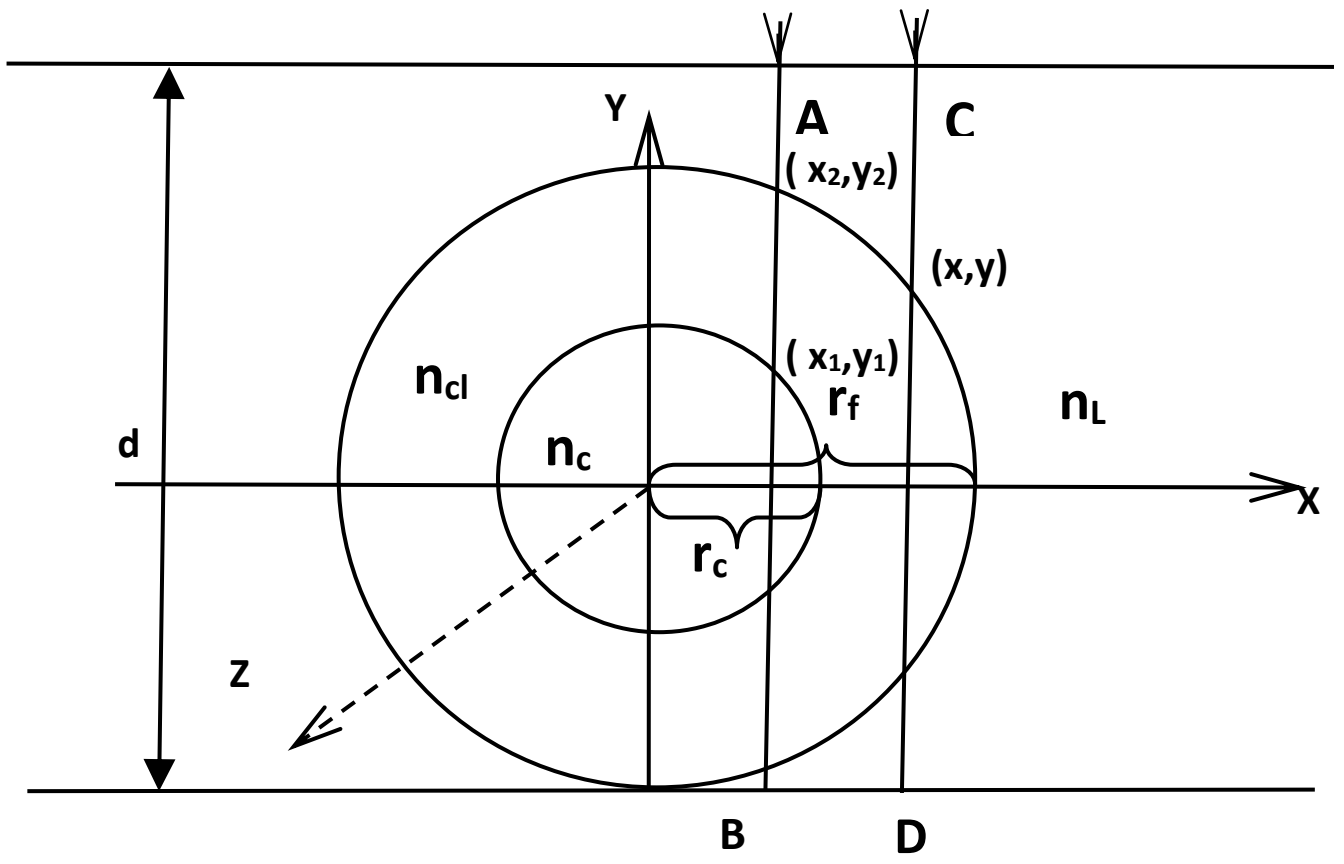

**Fig.1 Circular cross section of an optical fiber immersed in liquid silvered wedge at thickness d.**

Multiple beam interference (liquid wedge Fizeau interferometer) has been used to show the shape of interference fringes inside the fiber material. Light rays have been passed into three regions: liquid region, cladding region and core cladding region as shown in figure 1.

The four previous parameters and the liquid refractive index can form a certain shape for the interference fringes through the fiber according to the relation [14]:

$$\frac{\lambda}{4}\frac{\delta Z}{Z_l} = (n_{cl} - n_l)(r_f{}^2 - x^2)^{\frac{1}{2}} + (n_c - n_{cl})(r_c{}^2 - x^2)^{\frac{1}{2}} \qquad (1)$$

Where $Z_l$ is the separation between interference fringes inside the liquid region and $\delta Z$ the fringe shift inside the core region $0 \leq x \leq r_C$ .

## 2.2. Optical path in elliptical fiber.

The fiber material may suffer from some vertical compression when it is inserted inside the wedge of the interferometer especially, in case of polymer fiber. This compression will change the circular cross section to elliptical cross section as shown in figure2.So that equation (1) has to be modified. The horizontal radius will be elongated, the core radius $r_c$ changes to $a_c$ and the fiber radius $r_f$ changes to $a_f$. Also, the vertical radius will be compressed such that the core radius $r_c$ changes to $b_c$ and the fiber radius $r_f$ changes to $b_f$.

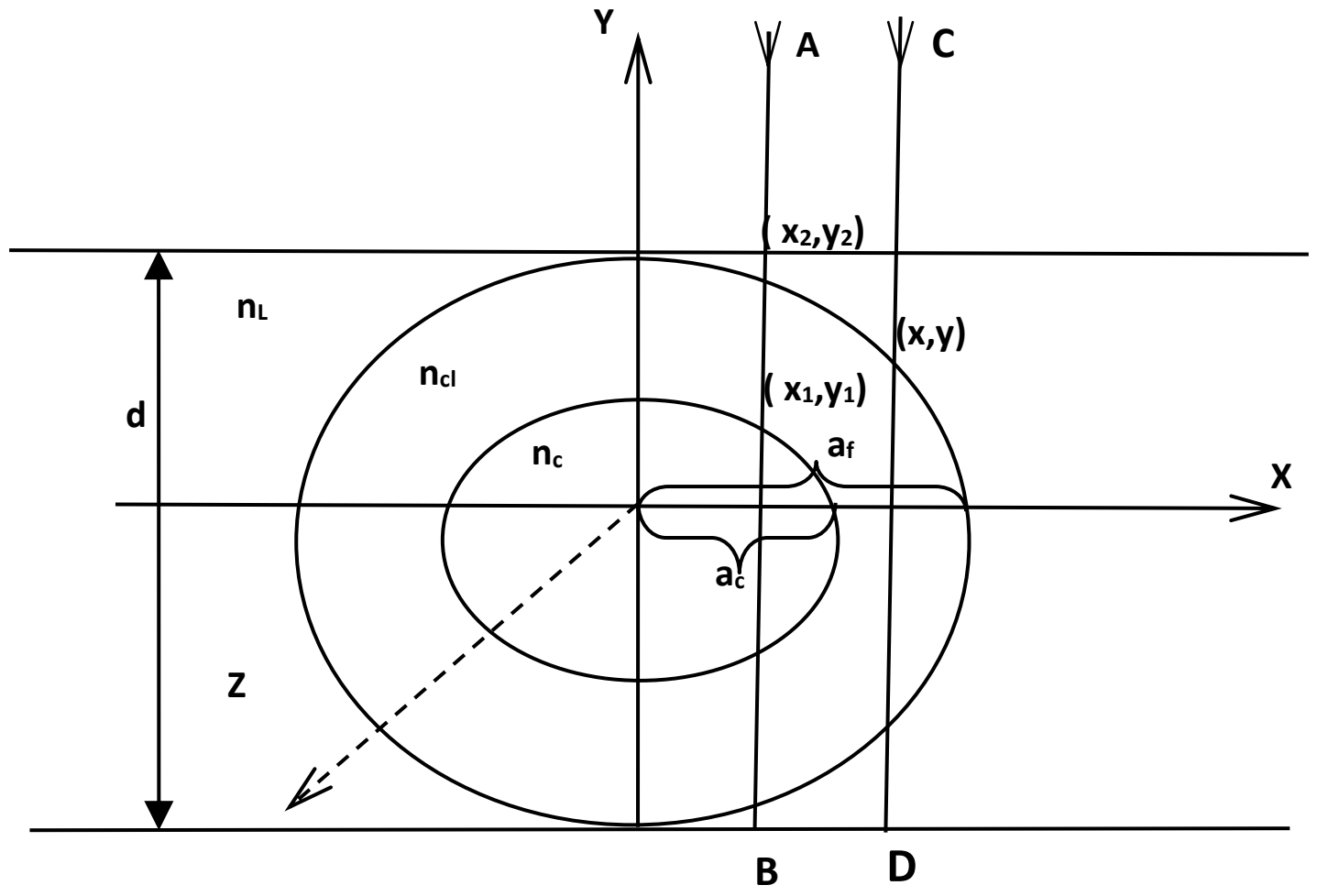

**Fig.2 Elliptical cross section of optical fiber immersed in liquid wedge interferometer at thickness d.**

The equation of elliptical shape $\frac{x^2}{a^2}+\frac{y^2}{b^2}=1$ (2) Put $e=\frac{b_c}{a_c}=\frac{b_f}{a_f}$ (3)

After modifying equation 1 taking in our respect the vertical compression

$$\frac{\delta Z\lambda}{4e\boldsymbol{Z_l}} = 4e(a_f^2-x^2)^{\frac{1}{2}}(n_{cl}-n_L)+4e(a_c^2-x^2)^{\frac{1}{2}}(n_c-n_{cl}) \qquad (4)$$

at the center of the fiber $\boldsymbol{x=0}$

$$\frac{\boldsymbol{\delta Z\lambda}}{4e\boldsymbol{Z_l}} = \boldsymbol{a_f(n_{cl}-n_L)+a_c(n_c-n_{cl})} \qquad (5)$$

Subtract equation (4) from equation (5) we can get

$$\frac{\lambda}{4e\boldsymbol{Z_l}}\left(\boldsymbol{\delta Z}_{(x=0)}-\boldsymbol{\delta Z}_{(x)}\right) = \left(\boldsymbol{a_f}-\left(a_f^2-x^2\right)^{\frac{1}{2}}\right)(n_{cl}-n_L)+\left(\boldsymbol{a_c}-\left(a_c^2-x^2\right)^{\frac{1}{2}}\right)(n_c-n_{cl}) \qquad (6)$$

The cladding index can be calculated form the relation:

$$n_{cl}=\frac{\frac{\lambda}{4e\boldsymbol{Z_l}}\left(\boldsymbol{\delta Z}_{(x=0)}-\boldsymbol{\delta Z}_{(x)}\right)+n_L\left(\boldsymbol{a_f}-\left(a_f^2-x^2\right)^{\frac{1}{2}}\right)-n_c\left(\boldsymbol{a_c}-\left(a_c^2-x^2\right)^{\frac{1}{2}}\right)}{\left(\boldsymbol{a_f}-\boldsymbol{a_c}-\left(a_f^2-x^2\right)^{\frac{1}{2}}+\left(a_c^2-x^2\right)^{\frac{1}{2}}\right)} \qquad \textbf{(7)}$$

As shown in figure 3, the fringe shift at the core center $\boldsymbol{\delta Z}_{(\boldsymbol{x}=\mathbf{0})}$ and the fringe shift at any point x inside the core region $\boldsymbol{\delta Z}_{(\boldsymbol{x})}$ can be calculated from the interferogram.

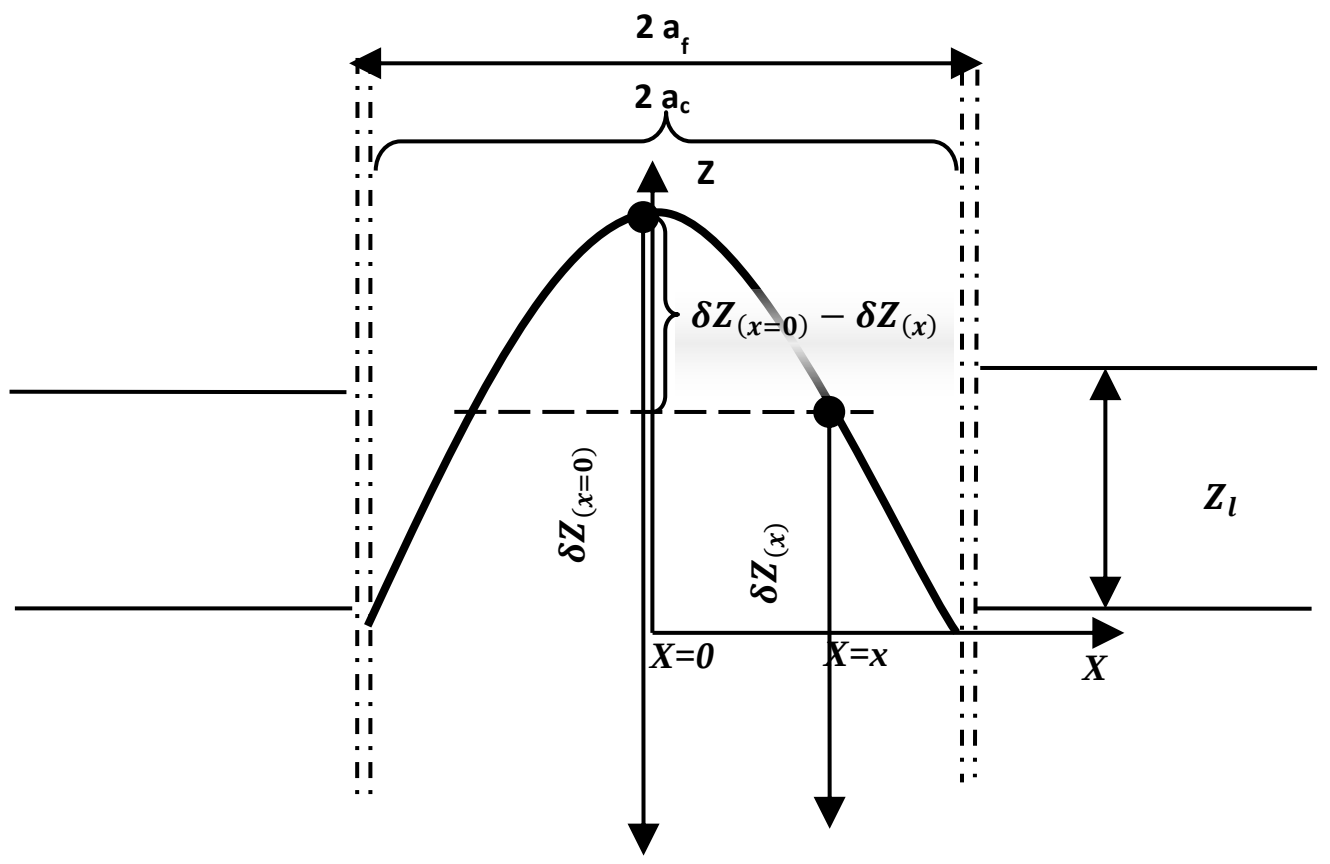

**Fig.3 the fringe shift inside the core region in case of large difference between the core and the cladding index**

# 3. Experimental Work.

## 3.1. Fiber Material.

The fiber sample has a core made of ploy methyl methacrylate (PMMA) and the cladding material is made of fluorinated polymer. The chemical structure is shown in figure4.

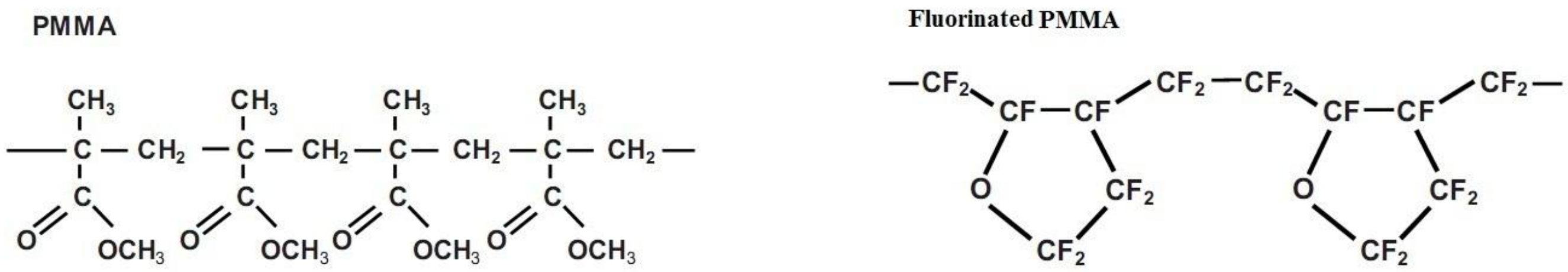

**Fig. 4 Chemical structure of PMMA and Fluorinated PMMA**

PMMA is produced from ethylene, hydrocyanic acid and methyl alcohol. It is resistant to water, dyes, diluted acids, petrol, mineral oil and turpentine oil. Also it is polymerized material that has amorphous structure. The refractive index of PMMA is 1.492 and its glass transition temperature ($T_g$) lies between +95°C and +125°C [3].

## 3.2. Plasma Irradiation.

The setup of pseudo-glow discharge plasma experiment consists of three main parts: Vacuum system, Pseudo-discharge tube and Power supply circuit. Continuous gas flow through the discharge vessel is maintained to sweep out the impurities from the system; the flow of the gas is controlled by a needle valve, and hence, adjusting the working gas pressure inside the gas tube.

The vessel of the cell is made of cylindrical Pyrex glass of 10 cm in diameter, 25 cm

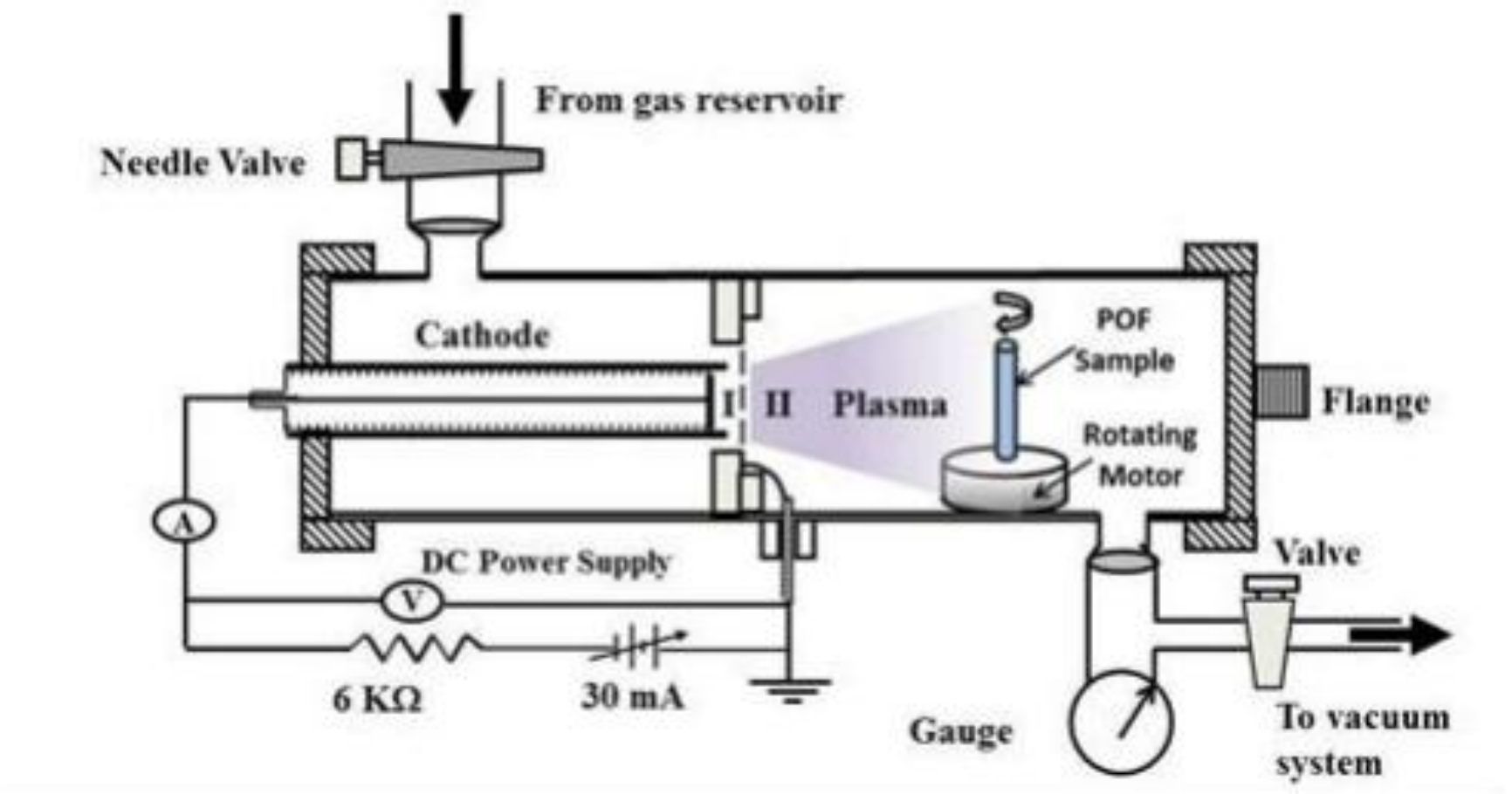

**Fig.5 Schematic diagram of discharge experiment arrangement.**

long. There are four ports sealed to the tube, the first one is used for the vacuum system pumping, the second one is employed for the gas inlet, and the two others, for diagnostics. The cathode is a copper rode of 2 cm diameter and 10 cm long, which is inserted in a Pyrex sleeve. The anode is a steel mesh 30 holes/inch$^2$ and placed at 4 mm from the cathode. The electrodes are fixed in the cell by two Perspex flanges as indicated in figure.5

The experimental tube will be evacuated and filled with the working gas at the required pressure P. Then, a sufficiently high voltage is applied across the two electrodes until a discharge current flows which means an electrical breakdown occurs. The expansion plasma is generated outside the two electrodes. This is clear since, the electrodes separation distance is less than or equal to the mean free path of the electrons. The electrons are accelerated in region I and diffused outside the two electrodes in region II.

The fiber sample has been hold in the plasma in region II and rotated over a small motor inserted on the plasma tube for different exposure time values and at certain pressure P= 0.4 torr and current intensity I=30mA and far from the anode by 5 cm.

## 3.3. Optical Setup

The fiber is immersed in a silvered liquid wedge interferometer and illuminated by a parallel monochromatic beam of light of wavelength $\lambda$ = 5461 A°.

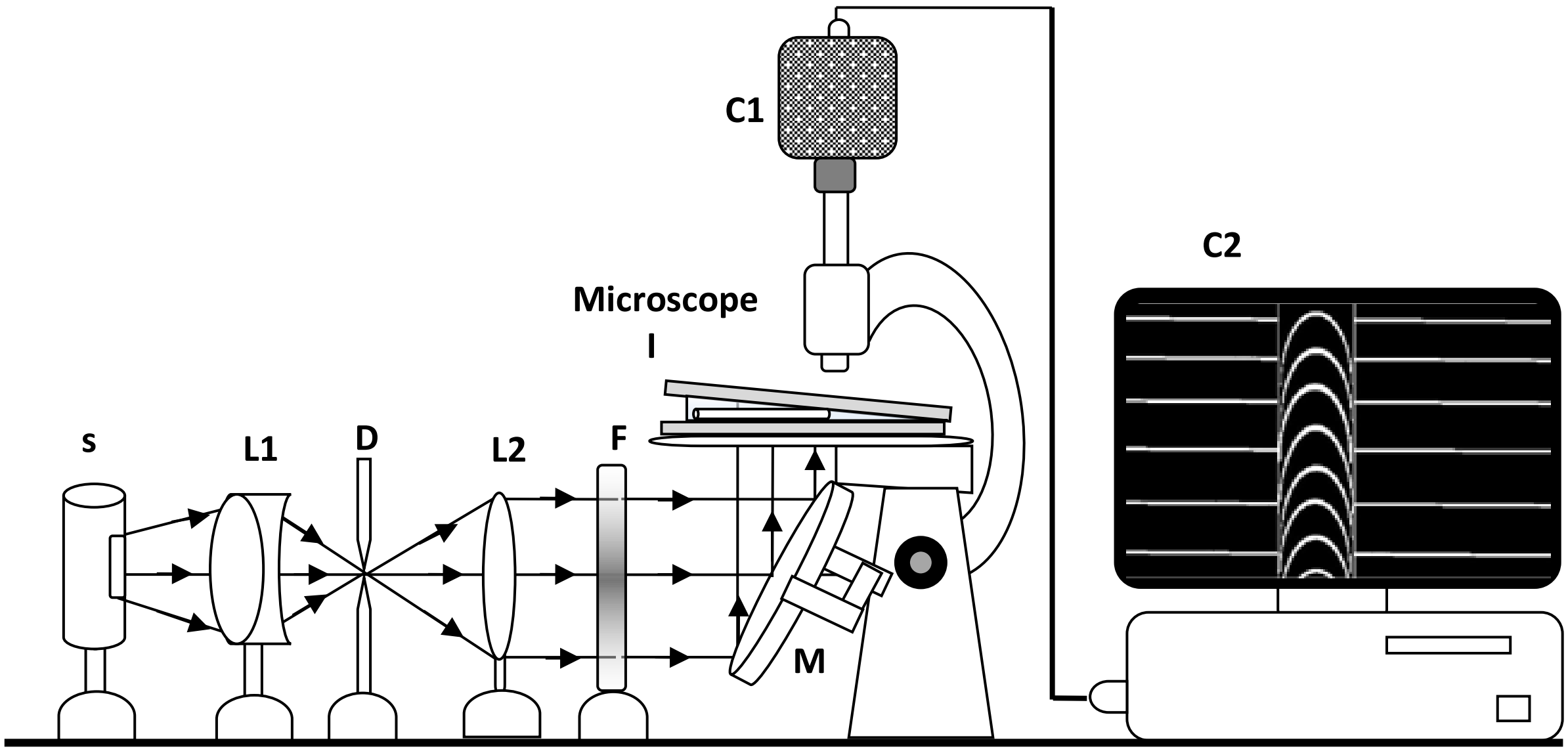

**Fig.6 Optical setup for producing multiple-beam Fizeau fringes in transmission. s, Mercury lamp; L1, condensing lens; D, iris diaphragm; L2, collimating lens; F, monochromatic Filter, M, reflecting mirror of microscope ; I, silvered liquid wedge interferometer; C1, to camera attached to microscope;C2, computer.**

The screws of the wedge interferometer have been adjusted to secure the interference fringes in the liquid region normal to the fiber axis. The shape and magnitude of the fringe shift across the fiber depend on the relative values of refractive indices of the immersion liquid, clad and core.

A photograph of interferogram shown in Fig. 6 is captured by a CCD camera, digitized by the frame grabber and saved in the computer's memory with Bmp format, the image is processed to remove noise and to make thinning for the interferogram. Then the required measurements on the fringes are performed. These processing steps enable us to locate the peak and will be mentioned in the next results.

# 4. Results.

## 4.1. Fringe Analysis.

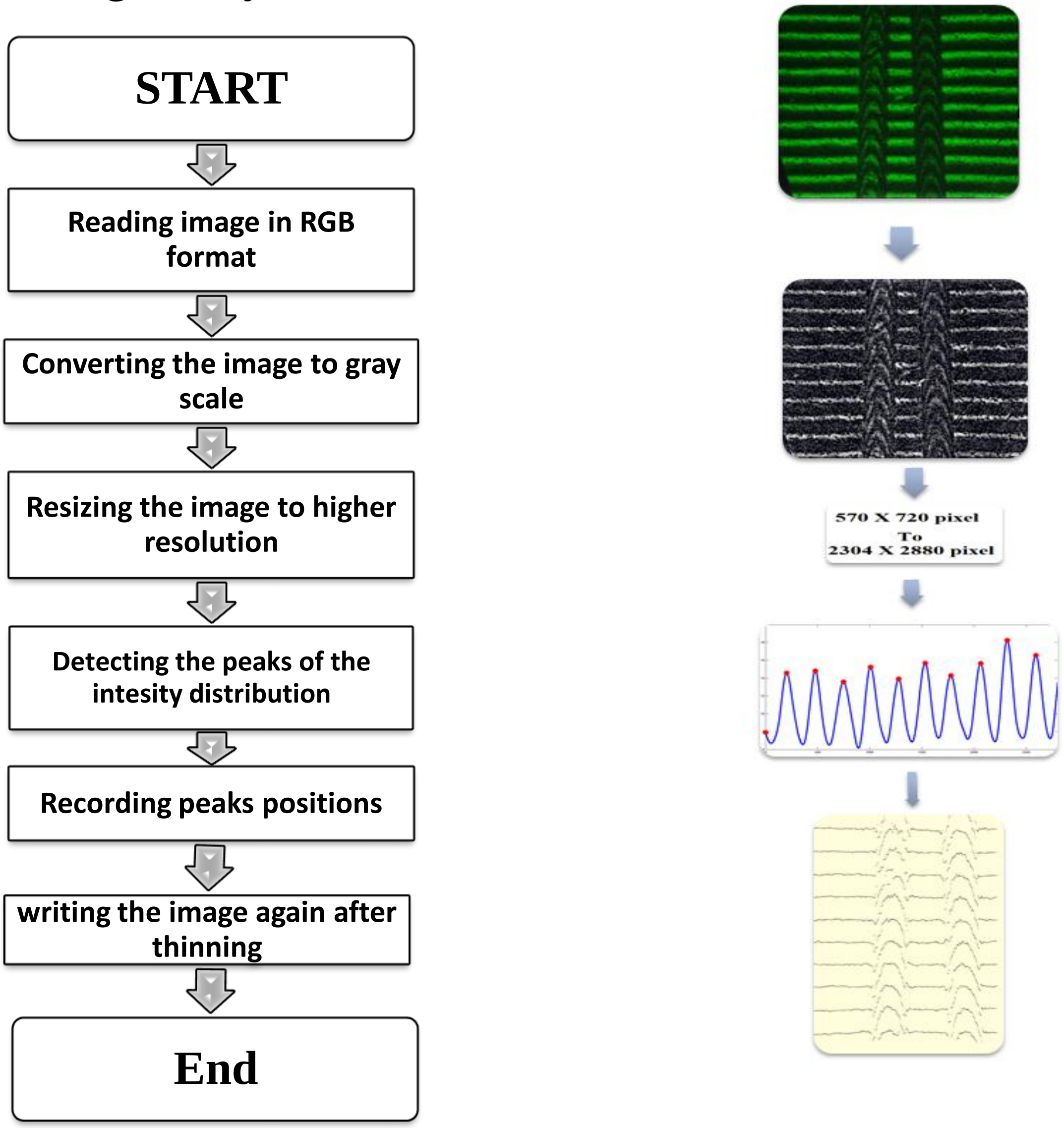

Fig.7 Flow chart of thinning process using Matlab program

Figure7 shows a flow chart of thinning process using Matlab program. Firstly the image is read by the program indicating the image location on the hard disc of the computer. Secondly the image is converted from RGB format to gray scale format and converted to a digital matrix of intensity values distributed on the image pixels. Thirdly the intensity distribution across specific line is drawn along the image. Across the drawing line, the peak locations on the image are determined and stored in a new matrix that has the same dimensions of the initial image. Fourthly plotting the new matrix gives us a graph containing thin lines without noise. Note that to get more details in the final image may be need some resizing to larger scale with higher image resolution.

## 4.2. Variation of cladding index.

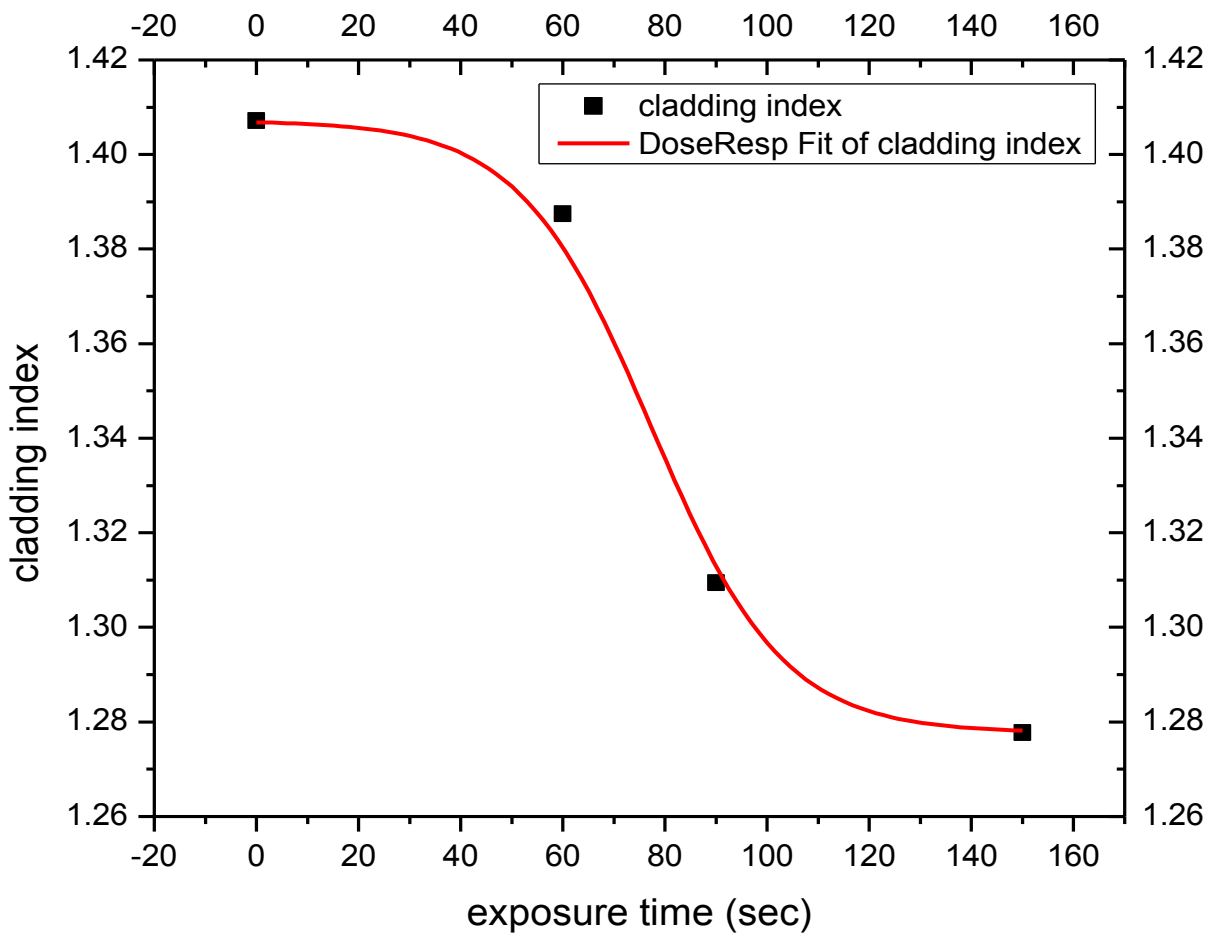

Fig.8 Variation of the cladding index with the exposure time (t)

Figure 8 shows the variation of the cladding refractive index with the exposure time. It is clear from a nonlinear fitting that applied to data points shown in equation 8; there is a decrease in the refractive index of the cladding region with increasing the exposure time. The cladding material is different to that of the core material so, it is expected that the plasma effect doesn't has the same behavior with the core.

$$n_{cl} = A_1 + \frac{(A_2 - A_1)}{1 + 10^{(\log(t_0) - t)p}} \quad (8)$$

| Parameters | value |
|---|---|
| $A_1$ | 1.27772 |
| $A_2$ | 1.40714 |
| $Log(t_0)$ | 77.30563 |
| p | -0.03367 |

**Table1. Fitting parameters of equation 8**

## 4.3. Variation of core index profile.

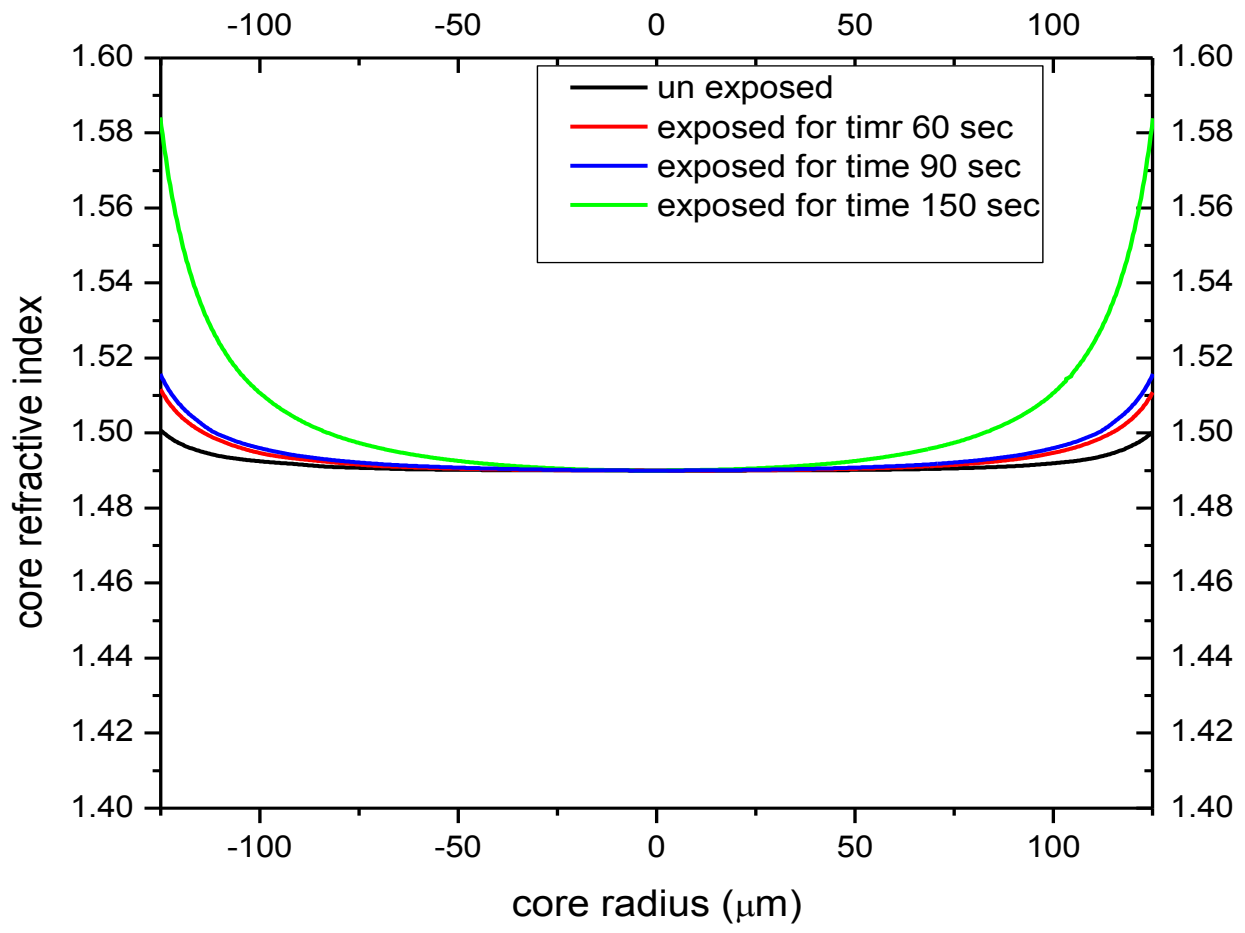

Fig.9. the core index profile variation with exposure time(t)

Figure 9 shows the variation of the core refractive index profile with the exposure time. It indicates that there is a regular increase in the peripheral core index with increasing the exposure time. This index variation decreases as we move toward the center of the fiber. It is an expected result since the plasma has a surface effect.

## 4.4. Skin depth due to plasma irradiation

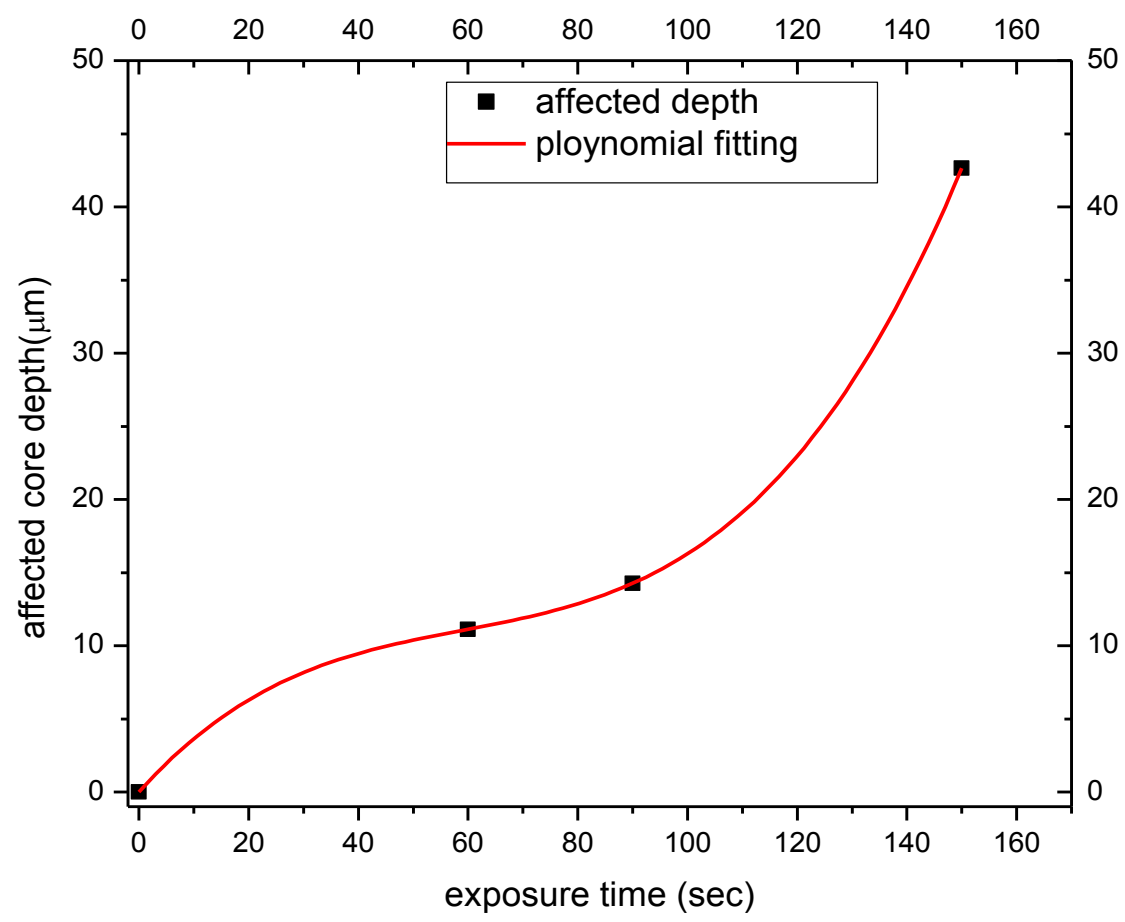

Fig.10 the affected depth in the core region with exposure time (t)

Figure 10 shows the affected depth in the core region with the exposure time. It is indicated that there is a regular increase in the peripheral core index with increasing the exposure time. A third-order fitting is applied to the data points that shows the increasing in the skin depth with the increasing in the exposure time .The plasma effect can change the refractive index in depth 42.6 μm in the core region for exposing time 150 sec.

$$\textbf{affected depth} = B_0 + B_1 t + B_2 t^2 + B_3 t^3 \quad (9)$$

| Parameter | Value(μm) |
|---|---|
| $B_0$ | $-1.065 \times 10^{-14}$ |
| $B_1$ | 0.41902 |
| $B_2$ | -0.00589 |
| $B_3$ | $3.32614 \times 10^{-5}$ |

**Table2. Fitting parameters of equation 9**

## 4.5. Numerical Aperture (N.A.)

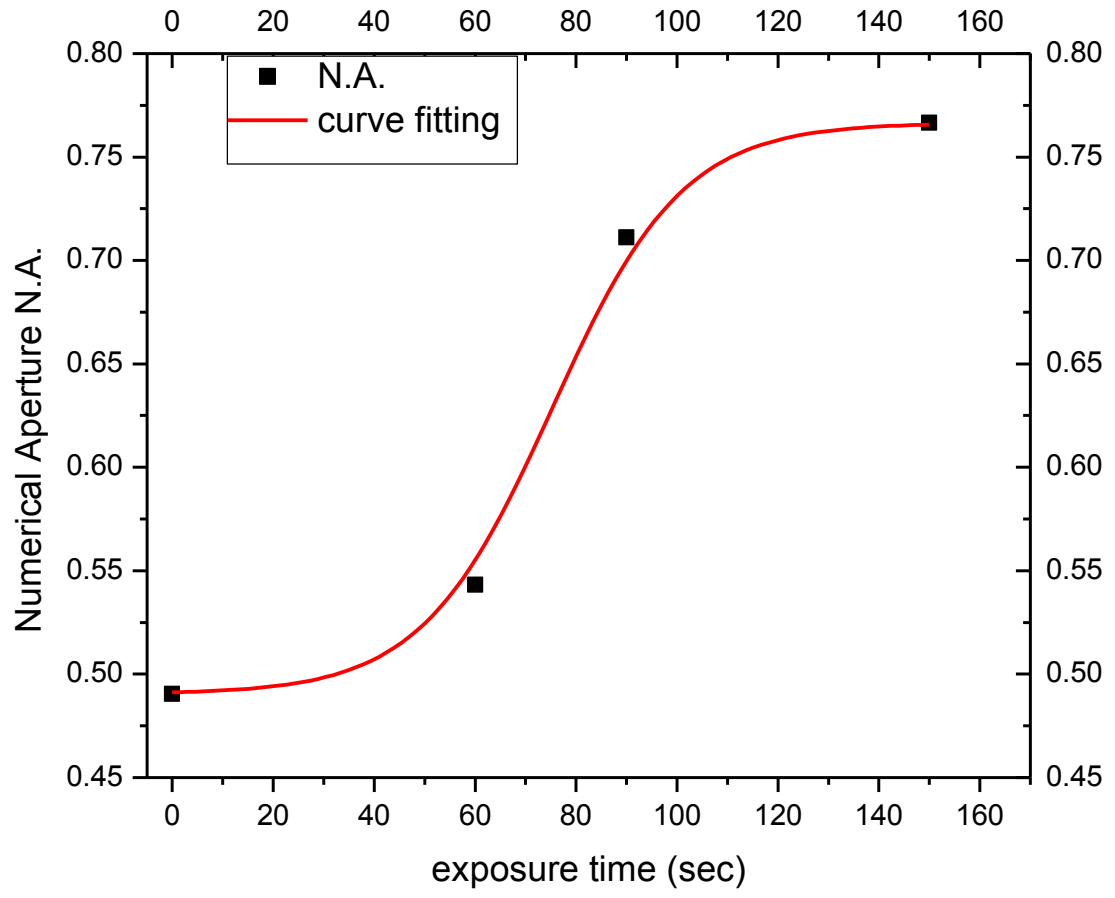

Fig.11. Numerical aperture (N.A.) variation with exposure time (t)

Figure 11 shows the numerical aperture variation with the exposure time. The numerical aperture is calculated according to equation 10, the center core refractive index isn't changed but the cladding refractive is changed .It is indicated that there is an increase in the numerical aperture with increasing the exposure time.

$$N.A. = \sqrt{n_c^2 - n_{cl}^2} \qquad (10)$$

$$N.A. = A_1 + \frac{(A_2 - A_1)}{1 + 10^{(\log(t_0) - t)p}} \quad (11)$$

| Parameters | value |
|---|---|
| $A_1$ | 0.49063 |
| $A_2$ | 0.76654 |
| $Log(t_0)$ | 75.24168 |
| p | $1.074453x10^{-5}$ |

**Table3. Fitting parameters of equation 11**

## 4.6. Maximum variation of core index

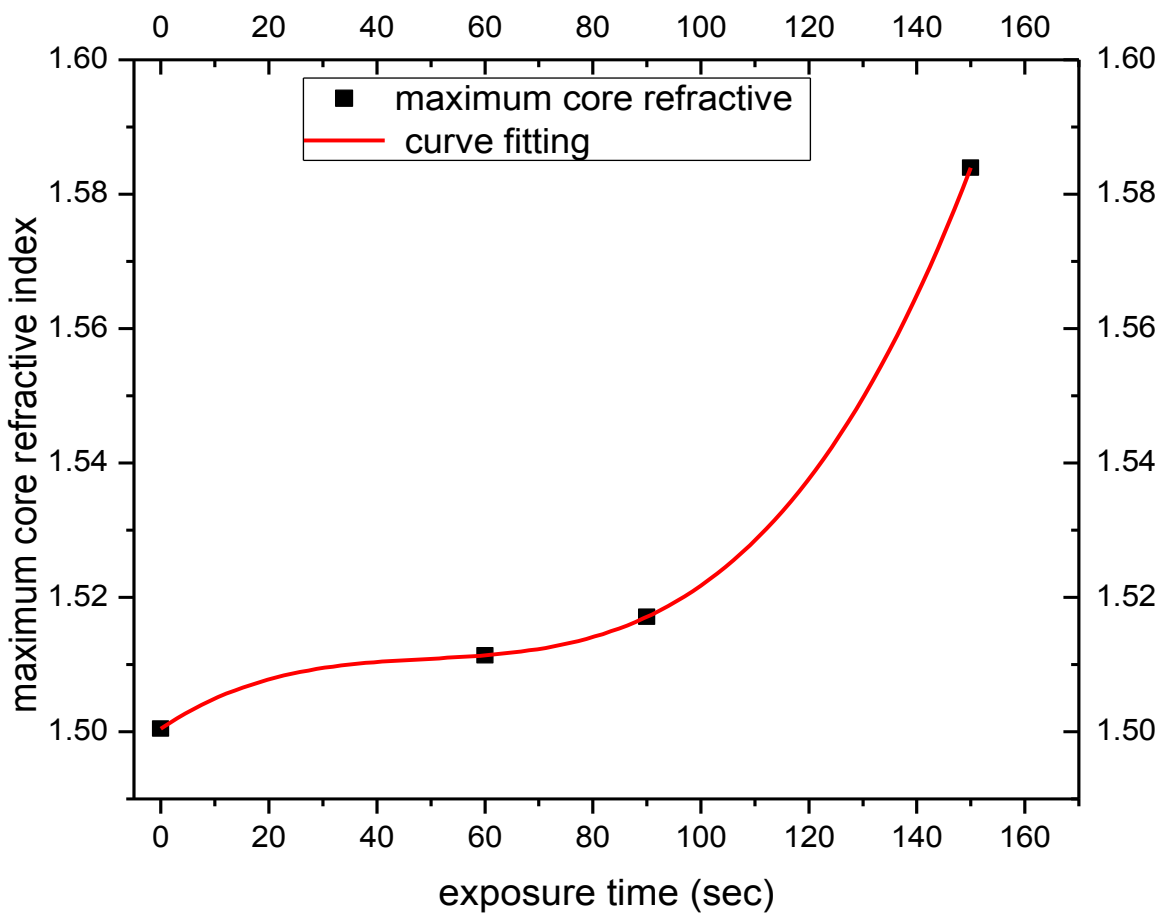

Fig.12 Maximum core refractive index variation with exposure time

Figure 12 shows the maximum core refractive index variation with the exposure time. The maximum value of the core refractive index located at the core-cladding interface. Also, the core refractive index at the core cladding interface increases with the increasing of the plasma exposure time.

$$\textbf{maximum core refractive} = B_0 + B_1 t + B_2 t^2 + B_3 t^3 \quad (12)$$

| Parameter | Value |
|---|---|
| $B_0$ | 1.5003 |
| $B_1$ | $5.42989x10^{-4}$ |
| $B_2$ | $-1.00741\ x10^{-5}$ |
| $B_3$ | $6.77654\ x\ 10^{-8}$ |

**Table4. Fitting parameters of equation 12**

Table.5 shows an overall view for plasma effect on calculated and measured optical parameters and other parameters are related to the signal propagation through the fiber core such as the normalized frequency (V-number) and number of propagating modes M.

| Exposure time (sec) | Cladding refractive index | Skin depth (μm) | Numerical Aperture | Maximum core refractive index | V-number | M (mode number) |
|---|---|---|---|---|---|---|
| 0 | 1.40714 | 0 | 0.49036 | 1.50043 | 676.67788 | 228946.47482 |
| 60 | 1.38748 | 11.12986 | 0.54314 | 1.51138 | 749.51254 | 280884.52203 |
| 90 | 1.30939 | 14.26887 | 0.71105 | 1.5171 | 981.23108 | 481407.21908 |
| 150 | 1.27772 | 42.6368 | 0.76654 | 1.58392 | 1057.79916 | 559469.52989 |

**Table.5 Optical parameters and with different values of cold plasma exposure time (t)**

## 5. FTIR. Spectroscopy

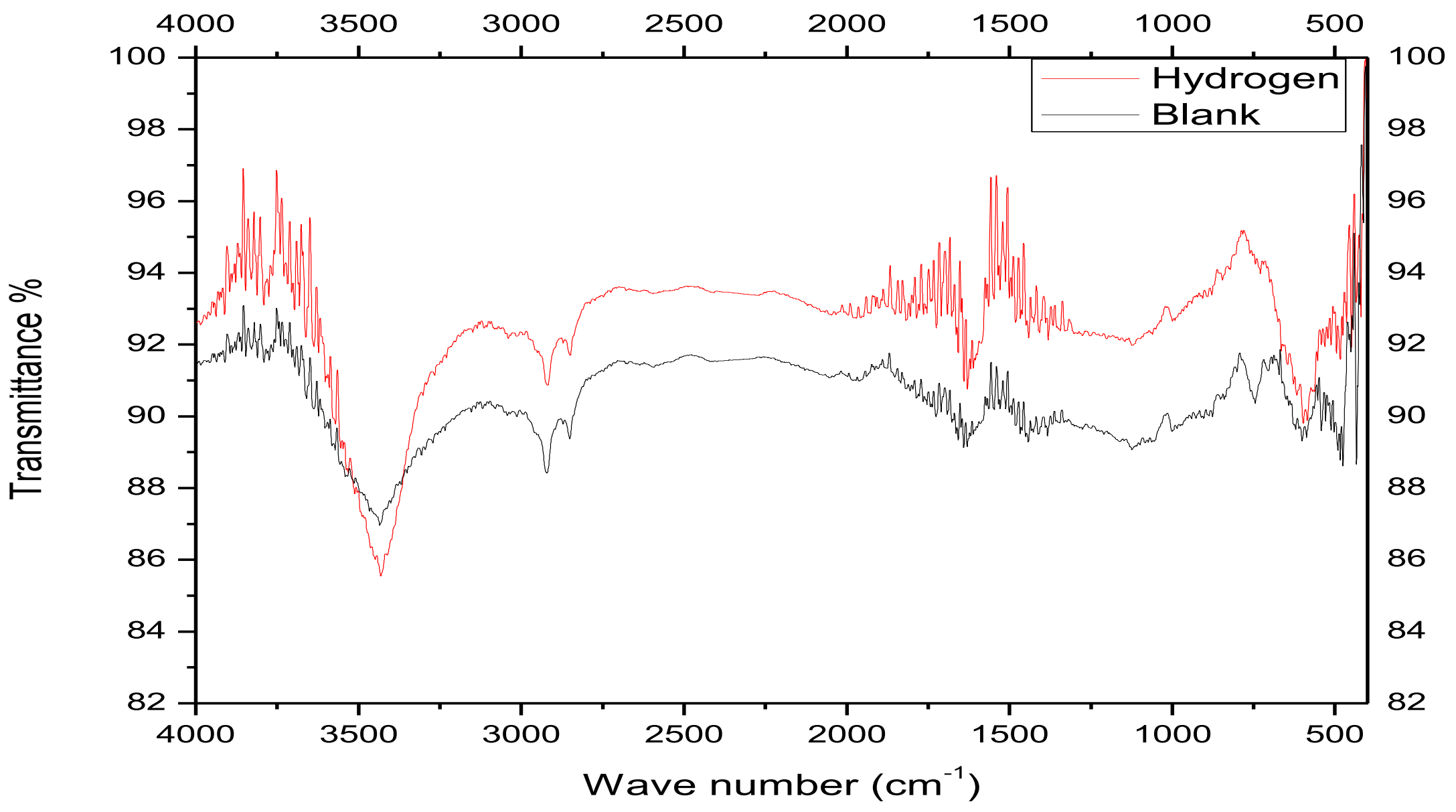

**Figure13.FTIR of plastic fiber unexposed plastic fiber sample (black line) and FTIR of exposed plastic fiber sample (red line)**

# 6. Conclusion

Plasma treatment has been applied on step index plastic optical fiber samples for different exposure times (0, 60, 90 and 150 sec.). The fiber samples have investigated using multiple beam interference (fizeau interferometer). We found that the plasma treatment has an observed effect on the optical parameters of the skin region of plastic optical fibers. With increasing the exposure time there is a regular increase in the peripheral core index. Also, the affected core depth and the maximum core refractive index are increased. The numerical aperture is increased due to the decrease of cladding refractive index with the increase of the exposure time. The increase of numerical aperture is preferable in communication field.

# References

(1) DeCusatis , Casimer M..; DeCusatis, Carolyn J.Sher.: Fiber Optic Essentials; California , Academic Press,2006.

(2) Murata, Hiroshi: Handbook of Optical Fibers and Cables; New York, Marcel Dekker, Inc,1996.

(3) Ziemann ,Olaf; Krauser, Jürgen; Zamzow, Peter E. ; Daum ,Werner : POF Handbook; Verlag Berlin Heidelberg, Springer,2008.

(4) Koike, Kotaro; Koike, Yasuhiro. Journal Of Lightwave Technology, VOL. 27, NO. 1, 2009.

(5) Park ,Soo-Jin; Cho ,Ki-Sook; Choi ,Choon-Gi. Journal of Colloid and Interface Science. 258 ,(2003), 424–426

(6) Barakat, N.; El-Hennawi, H. A.; El-Zaiat ,S. Y; Hassan,R. Pure Appl. Opt. 5,1996, 27–34

(7) El-Morsy ,M.A. ; Yatagai, T.; Hamza ,A.A.; Mabrouk ,M.A.; Sokkar,T.Z.N. Optics and Lasers in Engineering ,38, 2002,509–525

(8) Barakat, N.; El-Hennawi, H. A.; Abd. EL-Ghafar ,E.; EL-Ghandoor.H; Hassan,R;EL-Diasty,F.Optics communication ,191,2001,39-47

(9) Guorong, C.; Zheng, H.; Jun, X.; Jijian, C. J. Non-Cryst. Solids 2001, 288, 226–229.

(10) Luo, H.L.; Sheng, J.;Wan,Y.Z. Appl. Surf. Sci. 2007, 253, 5203–5207.

(11) Uhm, H.S.; Choi, E.H.; Choi, M.C. Appl. Phys. Lett. 2001, 79, 913–915.

(12) Grill, A. Cold Plasma in Materials Fabrication, Silver Spring, MD: IEEE Press, 1993.

(13) Kuptsov, a.h.; zhizhin, g.n. Handbook of Fourier transform Raman and infrared spectra of polymer, ELSEVIER SCIENCE PUBLISHERS B.V.,1998.

(14) Barakat,N. textile research journal.1971, 41